\documentstyle[prd,aps,twocolumn]{revtex}
\begin{document}
\draft
\def \d {{\rm d}}
\title{Boosted static multipole particles as sources of impulsive
gravitational waves}

\author{J. Podolsk\'y
 \thanks{E--mail: {\tt Podolsky@mbox.troja.mff.cuni.cz}}}
\address{Department of Theoretical Physics,\\
Faculty of Mathematics and Physics, Charles University,\\
V Hole\v{s}ovi\v{c}k\'ach 2, 18000 Prague 8, Czech Republic.}

\author{J. B. Griffiths
 \thanks{E--mail: {\tt J.B.Griffiths@Lboro.ac.uk}}}
\address{Department of Mathematical Sciences,\\
Loughborough University,\\
Loughborough, Leicestershire LE11 3TU, U.K.}

\date{\today}

\maketitle
\begin{abstract}
It is shown that the known solutions for nonexpanding impulsive
gravitational waves generated by null particles of arbitrary multipole
structure can be obtained by boosting the Weyl solutions describing static
sources with arbitrary multipole moments, at least in a Minkowski background.
We also discuss the possibility of boosting static sources in \hbox{(anti-)}
de~Sitter backgrounds, for which exact solutions are not known, to obtain the
known solutions for null multipole particles in these backgrounds. 
\end{abstract}


\narrowtext

\section{Introduction}

The purpose of this paper is to clarify the relationship between two classes
of exact solutions of Einstein's equations describing different kinds of
multipole particles. For static sources, Weyl's class of axially symmetric
vacuum solutions is well known \cite{R1}. The asymptotically flat solutions of
this class can be interpreted as external fields of axisymmetric sources with
arbitrary multipole structures (for a thorough review and interpretation see
\cite{R2} and the references therein). For the case of null particles with a
multipole structure, exact solutions and the impulsive gravitational waves
generated by them have recently been obtained \cite{R3},\cite{R4}. In these
solutions, the null particles and impulsive waves are taken to propagate in
background spaces of constant curvature: namely Minkowski, de~Sitter or
anti-de~Sitter spaces. The question now arises as to what are the relations
between these two types of solutions.

It may first be observed that, if the monopole solution is taken to be
Schwarzschild, the relation in this case is clear. As shown by Aichelburg and
Sexl \cite{R5}, the impulsive gravitational ({\sl pp}) wave generated by a
single null monopole particle can be obtained by boosting a Schwarzschild
black hole to the speed of light while its mass is reduced to zero in an
appropriate way.  Using a similar method, Hotta and Tanaka \cite{R6} have
boosted the Schwarzschild--de~Sitter solution to obtain the spherical
gravitational wave generated by a pair of null monopole particles in a
de~Sitter background. They have also described a similar solution in an
anti-de~Sitter background. Further details of the boosts and the geometry of
the non-expanding wave surfaces formed in these cases have been given
elsewhere~\cite{R7}.

It is thus clear that null monopole solutions can be regarded as the limits of
static monopole solutions boosted to the speed of light. It would be natural
to regard null multipole particles as the limit of static multipole particles
boosted in a similar way. For the case of a Minkowski background, we will show
that this is indeed the case. However, no explicit exact solutions are known
which describe static sources of any multipole structure in a background with a
non-zero cosmological constant. Such relations in a (anti-) de~Sitter
background must thus remain unresolved.

\section{Multipole particles in a Minkowski background}

We first consider static axisymmetric solutions with zero cosmological
constant. Following Quevedo \cite{R2}, we use the line element in Weyl
coordinates 
 \begin{equation}
 \d s^2=e^{2\psi}\d t^2 -e^{-2\psi} \big[e^{2\gamma}(\d\varrho^2+\d z^2)
+\varrho^2\d\varphi^2 \big], 
 \end{equation}
 where $\psi$ and $\gamma$ are functions of $\varrho$ and $z$ only. In this
form, exact asymptotically flat vacuum solutions describing the external
fields of sources with a multipole structure in a Minkowski background are
given by 
 \begin{eqnarray}
 \psi&=&\sum_{m=0}^\infty {a_m\over r^{m+1}}\,P_m(\cos\theta) \\
\gamma&=&\sum_{m,n=0}^\infty {(m+1)(n+1)\over m+n+2}
{a_ma_n\over r^{m+n+2}} \big( P_{m+1}P_{n+1}-P_mP_n\big). \nonumber
 \end{eqnarray}
 In this expansion $r=\sqrt{\varrho^2+z^2}$, $\cos\theta=z/r$ and $P_m$ are the
Legendre polynomials with argument $\cos\theta$. The sequence of arbitrary
constants $a_m$ determine the $m^{\rm th}$ multipole moments of the source, at
least in the Newtonian limit. However, in relativity, the definitions of
multipole moments are more complicated (for a full discussion see \cite{R2}).
In particular, it may be noted that the above ``monopole'' case in which all
$a_m$s vanish except $a_0$ is the Curzon--Chazy solution. As has been shown
elsewhere \cite{R8} the source at the origin in this case has directional
properties and thus a non-spherical structure. On the other hand, the
Schwarzschild solution which is the unique asymptotically flat spherically
symmetric exterior field for a monopole has the potential $\psi$ for a rod
when written in Weyl coordinates.

We now consider boosting the solutions (2) to the limit in which the speed of
the source approaches that of light. As will be clear below, in this limit it
is necessary to scale the constants $a_m$ to zero in an appropriate way. In
this case $\psi\ll1$, and the function $\gamma$ is of second order and may be
neglected. It may be noted that in this approximation we approach the
Newtonian limit, and in this case the differences between the Schwarzschild
and Curzon--Chazy solutions mentioned above also vanish. Working at this level
of approximation, the expansion (2) describes the multipole moments of a
source at $r=0$ in a physically meaningful sense.

Introducing standard cartesian coordinates $x=\varrho\cos\varphi$,
$y=\varrho\sin\varphi$, the line element (1) may be written in the approximate
form 
 \begin{eqnarray}
 \d s^2 &=&\d t^2-\d x^2-\d y^2-\d z^2 \nonumber\\
&& \qquad +2\psi(\d t^2+\d x^2+\d y^2+\d z^2).
 \end{eqnarray}
 First we boost this solution in a direction orthogonal to the axis of
symmetry which, without loss of generality, may be chosen as the $x$
direction: 
 $$ t={\tilde t+v\tilde x\over\sqrt{1-v^2}}, \qquad 
x={\tilde x+v\tilde t\over\sqrt{1-v^2}}. $$ 
 In the limit as $v\to1$ this yields 
 $$ \d s^2 =\d\tilde t^2-\d\tilde x^2-\d y^2-\d z^2 
+4(\d\tilde t+\d\tilde x)^2 \lim_{v\to1}{\psi\over1-v^2}. $$ 
 To evaluate this limit, we use the identity (employed elsewhere \cite{R6},
\cite{R7})
 \begin{equation}
\lim_{v\to1} {1\over\sqrt{1-v^2}}g(x) =\delta(\tilde t+\tilde x)
\int_{-\infty}^\infty g(x)\d x 
 \end{equation}
 which is valid in the distributional sense. It is thus necessary to rescale
the parameters $a_m$ to zero in the same way for each $m$,  such that \ 
$4a_m/\sqrt{1-v^2}=c_m$ \ where $c_m$ are a new set of constants which
characterise the multipole moments of the boosted source. The result is an
impulsive {\sl pp}-wave metric given by 
 \begin{eqnarray}
 \d s^2 &=&\d\tilde t^2-\d\tilde x^2-\d y^2-\d z^2 \nonumber\\
&& \qquad +H(y,z)\delta(\tilde t+\tilde x)(\d\tilde t+\d\tilde x)^2 
 \end{eqnarray}
 where 
 \begin{eqnarray}
 H&=&\sum_m H_m \\
&=&\sum_m c_m\int_{-\infty}^\infty {1\over(\rho^2+x^2)^{(m+1)/2}}\, 
P_m\!\left({z\over\sqrt{\rho^2+x^2}} \right) \d x \nonumber
 \end{eqnarray}
 where $\rho^2=y^2+z^2$.

We observe that, for the simplest case $c_0\ne0$, $c_m=0$ for $m\ge1$ which
corresponds to the boosted Curzon--Chazy solution, the integral diverges.
However, the divergence can be removed by first making the transformation 
 $$ \tilde t-\tilde x \to\tilde t-\tilde x +c_0\lim_{v\to1} 
\log\left(\tilde x+v\tilde t-\sqrt{(\tilde x+v\tilde t)^2+1-v^2}\right) $$ 
 which gives
 \begin{eqnarray}
	H_0&=&c_0\int_{-\infty}^\infty \left[ (\rho^2+x^2)^{-1/2}
-(1+x^2)^{-1/2} \right]\,\d x \nonumber\\
&=&-2c_0 \log\rho.\nonumber
 \end{eqnarray}
 This may be seen to be identical to the Aichelburg--Sexl solution \cite{R5}
which was originally obtained by boosting the Schwarzschild solution and
represents the impulsive gravitational wave generated by a single null
(monopole) particle.

For the higher multipole components each term may be considered separately.
For arbitrary $m\ge1$, we require to evaluate 
 $$ H_m = c_m\int_{-\infty}^\infty
(\rho^2+x^2)^{-(m+1)/2}\, P_m\!\left(z(\rho^2+x^2)^{-1/2}\right) \d x. $$ 
 Using the standard expression for the expansion of Legendre polynomials we
obtain
 \begin{eqnarray}
  H_m &=&c_m \sum_{k=0}^N {(-1)^k(2m-2k)!\over2^mk!(m-k)!(m-2k)!}\, z^{m-2k}
\nonumber\\
&&\qquad \times\int_{-\infty}^\infty (\rho^2+x^2)^{k-m-{1\over2}} \d x 
\nonumber
 \end{eqnarray}
 where $N={m-1\over2}$ if $m$ is odd, and $N={m\over2}$ if $m$ is even. Then,
using 
 $$ \int_{-\infty}^\infty (\rho^2+x^2)^{-p-{1\over2}} \d x 
={2^{2p}\over p}{(p!)^2\over(2p)!}{1\over\rho^{2p}} \qquad {\rm for} \quad
p=1,2,3\dots $$ 
 and putting $\cos\phi=z/\rho$ we get 
 $$ H_m(\rho,\phi) ={c_m\over\rho^m}\, \sum_{k=0}^N
{(-1)^k(m-k-1)!\over2^{2k-m}k!(m-2k)!}\, (\cos\phi)^{m-2k}. $$ 
 Finally, using the identity 
 $$ \cos m\phi = \sum_{k=0}^N 
{(-1)^km(m-k-1)!\over2^{2k-m+1}k!(m-2k)!}\, (\cos\phi)^{m-2k} $$
 we obtain that 
 \begin{equation}
 H_m(\rho,\phi) ={2c_m\over m}\,{\cos m\phi\over\rho^m}. 
 \end{equation}
 This simple term can now be seen to be the $m^{\rm th}$ multipole component
of the exact vacuum solution describing an impulsive gravitational wave with a
single source of arbitrary multipole structure as described in \cite{R3}.

For the sake of completeness, we now consider boosting the metric (3) with
the initial static source (2) in the direction of the axis of symmetry,
namely in the original $z$ direction. Using similar steps to those above, we
obtain 
 \begin{equation}
 \d s^2 =\d\tilde t^2-\d x^2-\d y^2-\d\tilde z^2 
+G(x,y)\delta(\tilde t+\tilde z)(\d\tilde t+\d\tilde z)^2,
 \end{equation}
 with 
 \begin{eqnarray}
 G&=&\sum_m G_m \nonumber\\
&=&\sum_m c_m\int_{-\infty}^\infty
{1\over(\rho^2+z^2)^{(m+1)/2}}\, P_m\!\left({z\over\sqrt{\rho^2+z^2}}\right) 
\d z \nonumber
 \end{eqnarray}
 where now $\rho^2=x^2+y^2$. For the simplest case, $c_0\ne0$, $c_m=0$ for
$m\ge1$, this is identical to the previous case ($G_0=H_0$) which yields the
Aichelburg--Sexl solution \cite{R5}. It is also obvious that $G_m=0$ for any
odd value of $m$. However, it can also be shown that $G_m=0$ for any $m\ge1$.
This can be observed by substituting $\xi=z/\sqrt{\rho^2+z^2}$ so that
 \begin{eqnarray}
 G_m&=& {c_m\over\rho^m} \int_{-1}^1 (1-\xi^2)^{m/2-1}P_m(\xi)\,\d\xi
\nonumber\\
 &=&{\pi c_m\over\rho^m} {\big(\Gamma({m\over2})\big)^2
\over\Gamma(m+{1\over2}) \Gamma({m\over2}+1) \Gamma(1-{m\over2})\Gamma(0)}
=0, \nonumber
 \end{eqnarray}
 which vanishes since $\Gamma(n)$ diverges at $n=0,-1,-2,-3,\dots$. This is
exactly in accord with intuition since the higher multipoles can be considered
in the Newtonian limit as mass distributions (of zero total mass) along the
$z$ axis. With a Lorentz contraction in the $z$ direction the effects of these
distributions will vanish leaving only the monopole term~$G_0$. 

\section{Multipole particles in (anti-) de~Sitter backgrounds}

In the case of a monopole boosted in the de~Sitter or anti-de~Sitter
backgrounds, exact solutions have already been described in \cite{R6} and
\cite{R7}. These have been obtained by boosting the
Schwarzschild--(anti-)de~Sitter solution using Lorentz transformations in the
five-dimensional representations of the (anti-) de~Sitter space-times. By a
different method, exact solutions describing impulsive gravitational waves in
the de~Sitter and anti-de~Sitter backgrounds with sources having an arbitrary
multipole structure have been given in \cite{R4}. It would be of interest to
see if these solutions could also be obtained by similarly boosting static
multipole sources. Unfortunately, no explicit static solutions are known in
space-times with a non-vanishing cosmological constant. However, we wish to
make some comments on the kind of solution that could be boosted to give the
known exact solutions for null multipole particles.

Of course, it is not possible to use an inverse process of ``slowing down'' a
null particle to obtain a solution for a static particle. We can therefore
only speculate on the form of the solution describing a static multipole
source in an (anti-) de~Sitter background. Nevertheless, there are some clues
at least to the linear approximation of the form of such a solution and we
will discuss these below.

It seems natural to consider the initial metric to be a perturbation of 
(anti-) de~Sitter space which reduces to the known form for the monopole case.
We therefore assume the line element in the form
 \begin{eqnarray}
 \d s^2 &=&\left(1-\epsilon{r^2\over a^2}\right)\d t^2 
-\left(1-\epsilon{r^2\over a^2}\right)^{-1}\d r^2 \nonumber \\
&&\hskip2cm -r^2(\d\vartheta^2+\sin^2\vartheta\d\varphi^2) \nonumber \\
&&\hskip2cm -\psi\left[ \d t^2+
\left(1-\epsilon{r^2\over a^2}\right)^{-2}\d r^2 \right],
 \end{eqnarray}
 where $a^2=3/|\Lambda|$ for a cosmological constant $\Lambda$, $\epsilon=1$
for a de~Sitter background ($\Lambda>0$), $\epsilon=-1$
for an anti-de~Sitter background ($\Lambda<0$), and $\psi$ is a function which
is independent of the coordinate $\varphi$ so that the metric (9) is axially
symmetric. Note that, as $\Lambda\to0$, the line element (9) reduces to a
perturbation of Minkowski space-time similar to~(3). For $\psi=2M/r$ it
represents a monopole perturbation corresponding to the
Schwarzschild--\hbox{(anti-)}de~Sitter solution which was boosted to the
ultrarelativistic limit in \cite{R6} and \cite{R7} to yield exact solutions for
impulsive gravitational waves with null monopole sources.

When the perturbation function $\psi$ vanishes, the metric (9) reduces to the
line element of the (anti-) de~Sitter space-time in static coordinates. It is
convenient to express these de~Sitter or anti-de~Sitter backgrounds as a
four-dimensional hyperboloid 
 $$ {Z_0}^2 -{Z_1}^2 -{Z_2}^2 -{Z_3}^2 -\epsilon{Z_4}^2 =-\epsilon a^2 $$
 embedded in a five-dimensional Minkowski space-time
 $$ \d s_0^2= \d{Z_0}^2 -\d{Z_1}^2 -\d{Z_2}^2 -\d{Z_3}^2 -\epsilon\d{Z_4}^2 $$ 
 where the parametrization is given by 
 $$ Z_1=r\sin\vartheta\cos\varphi, \quad
 Z_3=r\sin\vartheta\sin\varphi, \quad
 Z_2=r\cos\vartheta  $$ 
 and for $\epsilon=1$: 
 $$ Z_0=\sqrt{a^2-r^2}\,\sinh(t/a), \quad
 Z_4=\pm\sqrt{a^2-r^2}\,\cosh(t/a), $$ 
 or for $\epsilon=-1$: 
 $$ Z_0=\sqrt{a^2+r^2}\,\sin(t/a), \quad
 Z_4=\sqrt{a^2+r^2}\,\cos(t/a).  $$ 
 Writing (9) as $\d s^2=\d s_0^2+\d s_1^2$, we may express the perturbation
as 
 \begin{eqnarray}
 \d s_1^2 &=&-a^2 \psi  \Bigg[
\left( {Z_4\d Z_0-Z_0\d Z_4\over{Z_4}^2-\epsilon{Z_0}^2} \right)^2 \nonumber\\
&&\hskip1.5cm +{a^2\over r^2}
\left( {Z_0\d Z_0-\epsilon Z_4\d Z_4\over{Z_4}^2-\epsilon{Z_0}^2} \right)^2
\Bigg], \nonumber
 \end{eqnarray}
 where $r=\sqrt{{Z_0}^2+\epsilon(a^2-{Z_4}^2)}$.

Performing a boost in the $Z_1$ direction, 
 $$ Z_0={\tilde Z_0+v\tilde Z_1\over\sqrt{1-v^2}}, \qquad
Z_1={\tilde Z_1+v\tilde Z_0\over\sqrt{1-v^2}},  $$
 the background is invariant. However, when boosting the source term $\psi$ as
above, we must rescale its multiplicative constants as
$4a_m/\sqrt{1-v^2}=c_m$. Then, using the identity (4), it can be shown that
$\d s_1^2$ takes the form 
 \begin{equation}
 $$  \d s_1^2 =-a^2H(Z_2,Z_4)\,\delta(\tilde Z_0+\tilde Z_1)\,
(\d\tilde Z_0+\d\tilde Z_1)^2 , 
 \end{equation}
 where, 
 \begin{equation}
 H(Z_2,Z_4) 
=\int_{-\infty}^\infty {\epsilon(a^2-{Z_4}^2){Z_4}^2+(a^2+{Z_4}^2)x^2 \over
\left({Z_4}^2-\epsilon x^2\right)^2}\, {\psi\over r^2}\, \d x 
 \end{equation}
 and we must make the substitutions 
 \begin{eqnarray}
 &&\cos\vartheta=Z_2/r, \qquad
r=\sqrt{x^2+\epsilon(a^2-{Z_4}^2)}, \nonumber\\ 
 &&\cos[h]^2(t/a)={Z_4}^2/({Z_4}^2-\epsilon x^2) \nonumber
 \end{eqnarray}
 in the expression for $\psi/r^2$, where $\cos(t/a)$ is used for the case
$\Lambda<0$ and $\cosh(t/a)$ for $\Lambda>0$ in the last expression. At
this point, we may observe that the structure of the metric perturbation (10)
is exactly that required to give the exact null multipole solutions obtained
in \cite{R4}. These solutions include impulsive gravitational waves located on
the null hypersurfaces given by ${Z_2}^2+{Z_3}^2+\epsilon{Z_4}^2=\epsilon a^2$,
which at any time is a two-sphere in a de~Sitter universe or a 2-dimensional
hyperboloid in an anti-de~Sitter universe as described in \cite{R7}. An
appropriate parametrization of this wave surface is given by 
 \begin{eqnarray}
  Z_2&=&a\sqrt{\epsilon(1-z^2)}\cos\phi, \nonumber\\ 
Z_3&=&a\sqrt{\epsilon(1-z^2)}\sin\phi, \nonumber\\
 Z_4&=&az, \nonumber
 \end{eqnarray}
 where $|z|\le1$ when $\epsilon=1$ and $|z|\ge1$ when $\epsilon=-1$. In terms
of these parameters, it was shown in \cite{R4} that a general family of vacuum
solutions can be expressed in the form 
 \begin{eqnarray}
 H(z,\phi)&=& \sum_m c_mH_m(z,\phi) \nonumber\\
&=& \sum_m c_mQ^m_1(z)\cos[m(\phi-\phi_m)], 
 \end{eqnarray}
 where $c_m$ and $\phi_m$ are real constants representing the arbitrary
amplitude and phase of each multipole component and $Q^m_1(z)$ are the
associated Legendre functions of the second kind.

Our purpose now is to try to relate the given multipole term $H_m(z,\phi)$ of
the known exact solution (12) to a possible perturbation term $\psi_m$ in
(9). Such a relation is given by the integral (11), but this of course is a
non-unique inverse problem.

In order to evaluate the integral (11) and express it in terms of the
parameters $z$ and $\phi$, it is convenient to make the substitution from $x$
to $\theta$ such that \ $x=a\sqrt{\epsilon(1-z^2)}\cot\theta$, \ after which
it takes the form 
 \begin{equation}
  H(z,\phi) =a[\epsilon(1-z^2)]^{3/2} \int_0^\pi
{z^2+\cos^2\theta\over(z^2-\cos^2\theta)^2} \, {\psi(r,\vartheta,t)\over r^2}
\,\d\theta, 
 \end{equation}
 in which the coordinates $r,\vartheta,t$ must be expressed using the
relations 
 \begin{eqnarray}
 &&r={a\sqrt{\epsilon(1-z^2)}\over\sin\theta}, \qquad
\cos\vartheta=\sin\theta\cos\phi, \nonumber \\
&&{\rm cos[h]}^2{t\over a}={z^2\sin^2\theta\over z^2-\cos^2\theta}. 
 \end{eqnarray}

First we observe that putting $\psi_0=a_0/r$ gives exactly the monopole case 
 $$ a^2H_0={1\over4}\int_0^\pi {z^2+\cos^2\theta\over(z^2-\cos^2\theta)^2}
\, \sin^3\theta \,\d\theta
={z\over2}\log\left|{1+z\over1-z}\right|-1 $$ 
 as described in \cite{R6} and \cite{R7}. Secondly, we observe that the correct
dependence of $H_m$ on $\phi$ required by (12) can be obtained if we assume
that $\psi_m$ is a polynomial of order $m$ in $\cos\vartheta$: 
 $$ \psi_m(r,\vartheta,t) =\sum_{n=0}^m \chi_n^{(m)}(r,t)\cos^n\vartheta $$ 
 for some suitable functions $\chi_n^{(m)}(r,t)$. However, it can be shown
that the most natural assumption (2), namely
$\psi_m=a_mr^{-m-1}P_m(\cos\vartheta)$, does {\it not} yield the required
result for $m\ge1$. Instead, for example, the function 
 \begin{eqnarray}
  \psi_1 &=&2\epsilon a_1{a^2r^2\over\sqrt{a^2-\epsilon r^2}}
\left({r^2\cos[h]^2(t/a)+3a^2\sin[h]^2(t/a) \over 
r^4\cos[h]^4(t/a)-a^4\sin[h]^4(t/a)} \right) \nonumber\\
 &&\hskip1.5cm \times\cos[h](t/a) \cos\vartheta 
 \end{eqnarray}
 does give the correct null dipole term 
 \begin{eqnarray}
 H_1&=&Q_1^1(z)\cos\phi \nonumber\\
&=&-\epsilon \sqrt{\epsilon(1-z^2)}
\left({1\over2}\log\left|{1+z\over1-z}\right|-{z\over z^2-1}\right) \cos\phi
\nonumber
 \end{eqnarray}
as given in \cite{R4}. This illustrates the fact that an explicit
$t$-dependence may be necessary in the linear approximation of the source term
in the static background coordinate system. This is unexpected, but may be
necessary to maintain the multipole structure in the everywhere curved (anti-)
de~Sitter background. However, the expression (15) may not be the unique
linear perturbation term for the non-null dipole. It also diverges in the
limit as $\Lambda\to0$.

\section{Conclusions}

For the situation in which the cosmological constant vanishes, we have
demonstrated the following results:

\begin{itemize}
\item Boosting the asymptotically flat Weyl solutions is an appropriate (but
not unique) way of finding exact solutions for null multipole particles. 

\item The solutions obtained in this way are exactly the impulsive 
{\sl pp}-wave solutions described elsewhere~\cite{R3}. 

\item Boosting the asymptotically flat Weyl solutions in the direction of the
axis of symmetry simply yields the Aichelburg--Sexl solution~\cite{R5} for a
null monopole particle. 
\end{itemize}

For the case of a non-zero cosmological constant, no exact solutions for
static multipole sources are known. However, solutions for null multipole
particles have been obtained~\cite{R4}. These describe nonexpanding impulsive
gravitational waves with null point sources. For the monopole case, the
solutions for null particles were obtained by boosting the
Schwarzschild--(anti-) de~Sitter solution \cite{R6}, \cite{R7}. The following
points have here been demonstrated: 

\begin{itemize}
\item A simple generalization of the Weyl solutions which describes the field
of a static multipole source in an asymptotically (anti-) de~Sitter space-time
cannot be obtained. 

\item The {\it structure} of the known solutions describing null multipole
particles in an \hbox{(anti-)} de~Sitter universe can be obtained by boosting
an appropriate (approximate) multipole solution. However, precise forms of the
wave amplitudes are difficult to find. 

\item An exact solution which represents a non-null multipole source in an
asymptotically (anti-) de~Sitter space-time may have to be non-static. 
\end{itemize}

\acknowledgments

This work was supported by a visiting fellowship from the Royal Society and,
in part, by the grant GACR-202/96/0206 of the Czech Republic and the grant
GAUK-230/96 of Charles University.


\begin{references}


\bibitem{R1} D. Kramer, H. Stephani, M. A. H. MacCallum and E. Herlt,  
{\sl Exact solutions of Einstein's field equations}, \S18.1 (Cambridge
University Press, 1980) 

\bibitem{R2} H. Quevedo,  Fortschr. Phys. {\bf 38}, 733 (1990).

\bibitem{R3} J. B. Griffiths and J. Podolsk\'y, Phys. Lett. A {\bf 236}, 8
(1997).

\bibitem{R4} J. Podolsk\'y and J. B. Griffiths, Class. Quantum Grav. {\bf 15},
453 (1998).

\bibitem{R5} P. C. Aichelburg and R. U. Sexl, Gen. Rel. Grav. {\bf 2}, 303
(1971).

\bibitem{R6} M. Hotta and M. Tanaka, Class. Quantum Grav. {\bf 10}, 307
(1993). 

\bibitem{R7} J. Podolsk\'y and J. B. Griffiths, Phys. Rev. D {\bf 56}, 4756
(1997).

\bibitem{R8} S. M. Scott and P. Szekeres, Gen. Rel. Grav. {\bf 18}, 557 (1986).


\end{references}
\end{document}